\documentclass[useAMS,usenatbib]{mn2e}
\usepackage{graphicx,subfigure,ram,natbib}

\usepackage[colorlinks,citecolor=blue]{hyperref}

\begin{document}

\title[A radio-loud quasar in the early Universe]{J0906$+$6930: a radio-loud quasar in the early Universe}

\author[Zhang et al.]{
Yingkang Zhang$^{1,2}$, 
Tao An$^{1,8}$\thanks{Email: antao@shao.ac.cn}, 
S\'{a}ndor Frey$^{3,1}$,
Krisztina \'{E}. Gab\'{a}nyi$^{3}$,
Zsolt Paragi$^{4}$, 
\newauthor
Leonid I. Gurvits$^{4,6}$,
Bong Won Sohn$^{5}$,
Taehyun Jung$^{5}$,
Motoki Kino$^{5,7}$,
Baoqiang Lao$^{1}$, 
\newauthor
Yang Lu$^{1}$,
and
Prashanth Mohan$^{1}$
\vspace{2mm}
\\ 
$^{1}$ Shanghai Astronomical Observatory, Chinese Academy of Sciences, 80 Nandan Road, 200030 Shanghai, China \\
$^{2}$ University of Chinese Academy of Sciences, 19A Yuquanlu, Beijing 100049, China \\
$^{3}$ Konkoly Observatory, MTA Research Centre for Astronomy and Earth Sciences, Konkoly Thege Mikl\'os \'ut 15-17, H-1121 Budapest, Hungary \\
$^{4}$ Joint Institute for VLBI ERIC, Postbus 2, 7990 AA Dwingeloo, the Netherlands \\
$^{5}$ Korea Astronomy and Space Science Institute, 776 Daedeokdae-ro, Yuseong-gu, Daejeon 305-348, Republic of Korea \\
$^{6}$ Department of Astrodynamics and Space Missions, Delft University of Technology, Kluyverweg 1, 2629 HS Delft, the Netherlands \\
$^{7}$ National Astronomical Observatory of Japan, 2-21-1 Osawa, Mitaka, Tokyo, 181-8588, Japan \\ 
$^{8}$ Key Laboratory of Radio Astronomy, Chinese Academy of Sciences, 210008 Nanjing, China 
}

\date{}

\maketitle

\begin{abstract}
Radio-loud high-redshift quasars (HRQs), although only a few of them are known to date, are crucial for the studies of the growth of supermassive black holes (SMBHs) and the evolution of active galactic nuclei (AGN) at early cosmological epochs. Radio jets offer direct evidence of SMBHs, and their radio structures can be studied with the highest angular resolution using Very Long Baseline Interferometry (VLBI). Here we report on the observations of three HRQs (J0131$-$0321, J0906$+$6930, J1026$+$2542) at $z > 5$ using the Korean VLBI Network and VLBI Exploration of Radio Astrometry Arrays (together known as KaVA) with the purpose of studying their pc-scale jet properties. The observations were carried out at 22 and 43 GHz in 2016 January among the first-batch open-use experiments of KaVA. The quasar J0906$+$6930 was detected at 22 GHz but not at 43 GHz. The other two sources were not detected and upper limits to their compact radio emission are given. Archival VLBI imaging data and single-dish 15-GHz monitoring light curve of J0906$+$6930 were also acquired as complementary information. 
J0906$+$6930 shows a moderate-level variability at 15 GHz. The radio image is characterized by a core--jet structure with a total detectable size of $\sim$5~pc in projection. The brightness temperature, $\gtrsim$1.9\,$\times$\,$10^{11}$~K, indicates relativistic beaming of the jet.  
The radio properties of J0906$+$6930 are consistent with a blazar. 
Follow-up VLBI observations will be helpful for determining its structural variation.
\end{abstract}

\begin{keywords}
techniques: interferometric -- galaxies: high-redshift -- radio continuum: galaxies -- quasars: individual: J0131$-$0321 -- quasars: individual: J0906$+$6930 -- quasars: individual: J1026$+$2542
\end{keywords}

\section{Introduction}

High-redshift quasars (HRQs) are of significant importance to cosmology and galaxy studies as they provide crucial information of the accretion process and growth of supermassive black holes (SMBHs) when they just became active \citep[e.g.,][]{2012Sci...337..544V,2015ApJ...804..148V}, and the evolution and feedback of active galactic nuclei (AGN) in the early Universe \citep[e.g.,][]{2012ARA&A..50..455F,2012A&A...537L...8C}. Since the first extremely high-redshift quasar was discovered at $z \sim 5.8$ \citep[][]{2000AJ....120.1167F}, about 100 quasars have been detected at $z > 5.6$ with the Sloan Digital Sky Survey \citep[SDSS;][]{2000AJ....120.1579Y} and many other optical surveys \citep[][and references therein]{2015AJ....149..188J,2016ApJS..227...11B,2016ApJ...833..222J}.

Among the HRQs, the radio-loud subsample constitutes an attractive group since the radio jet is directly related to the SMBH activity and their radio structures can be studied with the highest angular resolution via Very Long Baseline Interferometry (VLBI). 
Studies of the correlation between the extragalactic jet parameters and redshift offer a unique way of testing cosmological models and understanding the evolution of the Universe.
In particular, the compact jets of the HRQs can be used for cosmological tests by studying the apparent angular size--redshift \citep[e.g.,][]{1999A&A...342..378G}, and apparent proper motion--redshift relations \citep[e.g.,][]{1999NewAR..43..757K}. 
However, the number of currently detected radio counterparts of HRQs is still quite small. 
About 30 radio-loud AGN have been studied at $z > 4.5$ \citep[e.g.][]{2004ApJ...610L...9R,2011A&A...531L...5F,2013MNRAS.431.1314F,2015MNRAS.450L..57G,2016MNRAS.463.3260C}. 
Until now, there are no more than 9 quasars that have been imaged with VLBI on mas scales at redshift $z > 5$, and only 5 of them have extended (multi-component) radio structures \citep{2016MNRAS.463.3260C}, including two radio objects at $z > 6$ \citep{2008A&A...484L..39F,2011A&A...531L...5F}. 
Nearly half of the $z > 5$ HRQs are identified as blazars.
The beamed jet emission from blazars makes them more detectable than unbeamed AGN in flux-density-limited radio observations. 
\citet{2011MNRAS.416..216V} found an apparent abundance of highly beamed (blazar-type) radio sources over the non-beamed radio source population at $z > 3$. \citet{2016MNRAS.461L..21G} proposed an obscuring bubble model to explain this discrepancy, where the central region of the AGN is only unobscured when viewed close to the direction of the jet. VLBI studies of the high-redshift radio AGN can be used to derive the jet parameters (e.g., the Doppler boosting factor and Lorentz factor), thus to discern whether the jet properties of HRQs are intrinsically different from those at low redshifts, and to verify the blazar nature of a source by confirming the presence of highly Doppler-boosted radio emission.

In the context of elucidating the most compact jet structure,
we observed three high-redshift radio-loud quasars: J0131$-$0321 ($z = 5.18$), J0906$+$6930 ($z = 5.47$), and J1026$+$2542 ($z = 5.266$) at 43 and 22 GHz
with a combined VLBI network named KaVA consisting of the Korean VLBI Network (KVN) and VLBI Exploration of Radio Astrometry (VERA) array of Japan.
These three sources were selected as they are the brightest in radio among the $z > 5$ quasars known to date. 
The observations at 22 and 43 GHz, with the highest resolution of 0.5 milli-arcsecond (mas), allow us to study the most compact structure of these highest-redshift radio AGN at a hitherto unreachable regime of short millimetre wavelengths (in the source's rest frame). 
Moreover, as mentioned above, the core--jet morphology is rarely detected in high-redshift quasars until now, while J0906$+$6930 and J1026$+$2542 are the unique two having prominent jet structure \citep{2004ApJ...610L...9R,2015MNRAS.446.2921F}. VLBI data at a new epoch are an important supplement for exploring their (possible) jet proper motion, or offer a crucial reference point for follow-up VLBI imaging for proper motion measurement in the future. 
In the present paper, archival VLBI data and single-dish radio light curve of J0906$+$6930 are also used for a comprehensive study of the source properties.
Section \ref{sec2} gives the description of the KaVA observations and data analysis. In Section \ref{sec3}, the KaVA observational results are presented. Section \ref{sec4} discusses the radio properties of J0906$+$6930, and a summary is given in Section \ref{sec5}. Throughout this paper, we use the following cosmological parameters: $\Omega_{\rm m} = 0.27$, $\Omega_{\Lambda} = 0.73$, $H_{0} = $70 km\,s$^{-1}$\,Mpc$^{-1}$ \citep{2011ApJS..192...18K}. At this distance, 1 mas angular size corresponds to a projected size of 6.2 pc.

\section{Observations and data reduction}
\label{sec2}

The Korean VLBI Network \citep[KVN;][]{2003ASPC..306..373M} is the first VLBI array dedicated to the mm-wavelength radio observations in East Asia. KVN consists of three 21-m-diameter radio telescopes (in Seoul, Ulsan, and Jeju Island, Korea), and is operated by the Korea Astronomy and Space Science Institute (KASI). KVN supports simultaneous observations in four receiver bands at 22, 43, 86 and 129 GHz \citep{2013PASP..125..539H}.
The VLBI Exploration of Radio Astrometry \citep[VERA;][]{2003ASPC..306..367K} array is a Japanese VLBI array which is dedicated to high-precision VLBI astrometry, and is operated by the National Astronomical Observatory of Japan (NAOJ). The VERA consists of four 20-m-diameter radio telescopes located at Mizusawa, Iriki, Ogasawara, and Ishigaki-jima in Japan. VERA telescopes operate with dual-beam system enabling phase-referencing VLBI observations at 22/43 GHz.
In recent years, the combined KVN and VERA Array, called KaVA, is in operation since 2013. The KaVA consists of seven radio telescopes, and has a remarkably good $(u,v)$ coverage on both short and long baselines, which is important for imaging extended AGN jet structure \citep[][]{2015PKAS...30..637N}. 

Our KaVA observations were carried out in 2016 January as one of the first-batch open-use experiments of KaVA, thus can also be used to test the high-resolution imaging capability of KaVA. This project is one of the first applications of KaVA for imaging studies of quasars. 
The observations were performed on two consecutive days (2016 January 14 and 15) but divided into three segments (project codes: k15ta01a, k15ta01b, k15ta01c) for scheduling convenience  (Table~\ref{tab:obs}). 
All seven telescopes participated in these experiments. At some periods of time during the observations, certain individual antennas showed abnormally high system temperatures due to bad weather. The affected data were thus deleted in the subsequent analysis. 
The observations lasted for a total of 19~h. J0131$-$0321 was observed for 4~h at both frequencies. J1026$+$2542 and J0906$+$6930 were observed for 2~h at each frequency, respectively. The data were recorded in 16 intermediate frequency channels (IFs) in left circular polarization, resulting in a total bandwidth of 256~MHz and a sampling rate of 1024~Mbit\,s$^{-1}$. The correlation was carried out in the Korea--Japan Correlation Center (KJCC) at Daejeon, Korea \citep{2014AJ....147...77L}, with an integration time of 1.63~s. Except for one scan on a bright quasar NRAO 150 for the purpose of fringe finding at the beginning of each subsession, most of the remaining time was spent on the target sources.

 \begin{table}
 	\caption{Information about the KaVA observations}
	\setlength{\tabcolsep}{4pt}
 	\begin{tabular}{cccc}
 	\hline
 	Project code & $\nu$ (GHz) & Date & Target sources (on-source time) \\
 	\hline
 	k15ta01a  & 22 & 2016 Jan 14 & J0131$-$0321 (4 h)  \\
 	          &    &             & J0906$+$6930 (2 h)  \\
 	          &    &             & J1026$+$2542 (2 h)  \\
 	k15ta01b  & 43 & 2016 Jan 14 & J1026$+$2542 (2 h)  \\ 
 		      &    &             & J0906$+$6930 (2 h)  \\
 	k15ta01c  & 43 & 2016 Jan 15 & J0131$-$0321 (4 h)  \\
 	\hline    
    \end{tabular} \label{tab:obs}
 \end{table}

\begin{figure*}
\centering
	\includegraphics[width=0.95\textwidth]{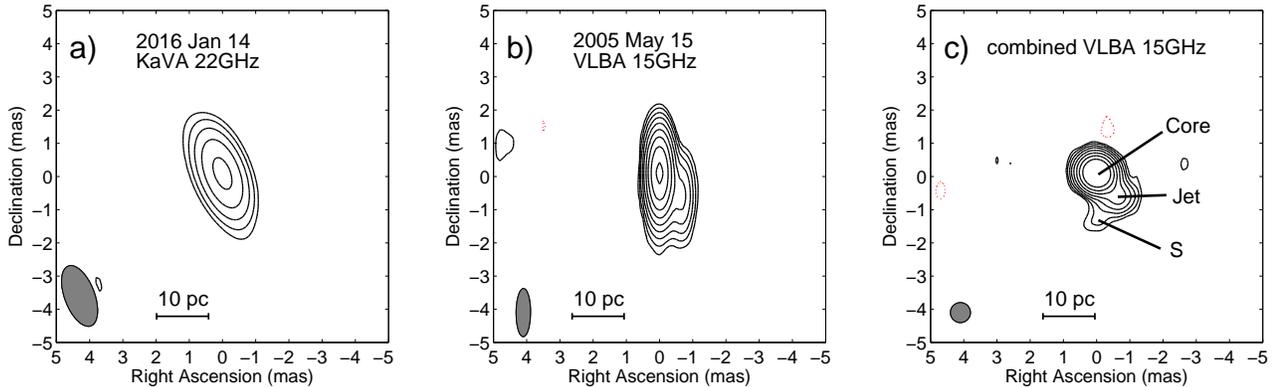}
	\caption{VLBI images of J0906$+$6930.
{\it a)} The KaVA image at 22~GHz. The peak intensity is 63.6~mJy\,beam$^{-1}$. The restoring beam is 1.9~mas\,$\times$\,0.9~mas at a position angle of $20\fdg7$, which is shown in the bottom-left corner of the image. The contours are 3 Jy beam$^{-1}$ $\times$ (1, 2, 4, 8, 16).
{\it b)} Naturally weighted 15-GHz VLBA image from the observations on 2005 May 15. The peak intensity is 116~mJy\,beam$^{-1}$. The restoring beam is 1.45 mas $\times$ 0.43 mas, at a position angle of $-0\fdg4$. The lowest contours represent intensity values of $\pm0.4$~mJy\,beam$^{-1}$ (3 times the rms noise in the image). The contours increase by a factor of 2.
{\it c)} The VLBA image created from the combined 15-GHz data. The peak intensity is 116.0~mJy\,beam$^{-1}$. The restoring beam is 0.6~mas\,$\times$\,0.6~mas. The lowest contour is at $\pm 3 \sigma$ (the rms noise level is 0.25~mJy\,beam $^{-1}$), and the contours increase by a factor of 2.
}
	\label{fig:VLBImap}
\end{figure*}

The correlated visibility data were imported into the NRAO Astronomical Image Processing System ({\sc AIPS}) software package \citep{2003ASSL..285..109G} in which the visibility amplitudes were calibrated using the system temperatures and antenna gains acquired by the stations. An amplitude calibration uncertainty of 15 per cent is estimated for the KaVA data \citep[see details in ][]{2014PASJ...66..103N}. Autocorrelation measurements were used to correct the cross-correlated visibility amplitudes. Note that for KaVA observations, a special amplitude correction factor of 1.3 is needed, as suggested by \citet{2015JKAS...48..229L}. The subsequent data reduction in AIPS followed the standard procedure described in \citet[][]{2015JKAS...48..229L}.  
The visibilities of the target sources were used to conduct fringe fitting with a solution interval of 2 min. In most of the observing time, the visibility phases show large scatter due to rapid tropospheric fluctuations. As a result, the signal-to-noise ratios (SNR) of the fringe-fitting solutions were mostly below 3. We could only find faint fringes in half of the 22-GHz data for the source J0906$+$6930. The 43-GHz data were too noisy to get clear fringes. The remaining two sources failed to show fringes either. We then imaged J0906$+$6930 at 22~GHz in the Caltech {\sc Difmap} package \citep{1994BAAS...26..987S}. As J0906$+$6930 is relatively weak, its total flux density measured with the Very Large Array (VLA) by \citet{2006AJ....132.1959R} is $\sim$80 mJy, and the source was marginally detected on KaVA baselines at 22~GHz, no self-calibration was attempted as it might introduce a spurious source at the phase center or increase the flux density abnormally \citep{2008A&A...480..289M}. Therefore we only performed CLEANing in order to reduce  the image noise.

In order to supplement the new KaVA data, we also re-analyzed archival Very Long Baseline Array (VLBA) data of J0906$+$6930 taken at 8.4 GHz (project code: BC196) and at 15~GHz (project codes: BR093, BG154). Table \ref{tab:obsVLBA} summarizes the VLBA observations. All ten antennas participated in the VLBA observations in 2011 June, 2004 Feb and 2005 May. Eight telescopes (all VLBA stations but St. Croix and Pie Town) participated in the 2005 March experiment. The data analysis followed the standard VLBA data reduction procedure \citep{1995ASPC...82..227D}. 
After calibration, we combined the three 15-GHz datasets in {\sc AIPS} to make a final image. The combined data set has a nearly circular ({\it u,v}) coverage and enables us to obtain the image shown in Fig. \ref{fig:VLBImap}-c.

\section{Results and comments on individual sources}
\label{sec3}
 
 \begin{table}
	\caption{Parameters of the KaVA observations}
 \begin{tabular}{ccccc}
 	\hline
	 Source name & $z$ & Frequency & Flux density & Restoring beam       \\
	             &	 & 	(GHz)	 &     (mJy) 	& (mas$^{2}$)      \\
	\hline
J0906$+$6930	 & 5.47 &	22	& $74\pm11$          & $1.90 \times 0.91$  \\	
				 &	    &	43  &  $<$57 (3$\sigma$) & $0.78 \times 0.48$  \\
J0131$-$0321	 & 5.18	& 	22	&  $<$53 (3$\sigma$) & $1.30 \times 0.99$  \\ 
				 &	    &	43 	&  $<$43 (3$\sigma$) & $0.90 \times 0.52$  \\
J1026$+$2542	 & 5.27	&  	22	&  $<$29 (3$\sigma$) & $1.10 \times 1.10$  \\
				 &	    &	43	&  $<$54 (3$\sigma$) & $0.73 \times 0.47$  \\
	\hline
 \end{tabular} \label{tab:flux}  
 \end{table}
 
 \begin{table*}
 	\caption{Summary of the archival VLBA datasets}
	\setlength{\tabcolsep}{4pt}
 	\begin{tabular}{cccccc}
 	\hline
 	Project code & $\nu$ (GHz) & Date & Bandwidth (MHz) & On-source time & Restoring beam \\
 	\hline
 	BC196   & 8.4 & 2011 Jun 03 & 128 &  7 min & 2.24 mas $\times$ 0.91 mas, PA = 20.2$\degr$ \\
 	BR093$^a$&15.4 & 2004 Feb 27 &  16 & 75 min & 1.46 mas $\times$ 0.47 mas, PA = 0.8$\degr$ \\ 
 	BG154   &14.4 & 2005 Mar 22 &  64 & 40 min & 1.60 mas $\times$ 0.51 mas, PA = 77.6$\degr$\\
	 	    &15.4 & 2005 May 15 &  16 &120 min & 1.45 mas $\times$ 0.43 mas, PA = $-0.5\degr$ \\
 	\hline    
    \end{tabular} \label{tab:obsVLBA} \\
$^a$ published in \citet{2004ApJ...610L...9R}
 \end{table*}

 \begin{table*}
 	\caption{Fitted parameters of the emission structure, based on VLBI observations of J0906$+$6930}
	\setlength{\tabcolsep}{4pt}
 	\begin{tabular}{ccccccc}
 	\hline
	Date & 	Telescope & $\nu$ (GHz) & Comp. & $S$ (mJy) & $\theta$ (mas) & $T_{\rm b}$ ($10^{11}$\,K)\\
 	\hline
 	2004 Feb 27 & VLBA   & 15.4& Core & $132 \pm 9$ & 0.13  & 2.6 \\ 
 	            &        &     &  Jet & $7.4 \pm 0.5$ & 0.10  & \\ 
    2005 Mar 22 & VLBA   & 14.4& Core & $136 \pm 9$ & 0.19  & 1.4\\
                &        &     &  Jet & $5.7 \pm 0.4$ & 0.10 &  \\
    2005 May 15 & VLBA   & 15.4& Core & $124 \pm 8$ & 0.15  & 1.9 \\
                &        &     &  Jet & $9.2 \pm 0.6$ & 0.10 & \\
	2011 Jun 03 & VLBA   & 8.4 & Core & $167 \pm 11$ & 0.33 & 1.8 \\
 	2016 Jan 14 & KaVA   &22.2 & Core & $74 \pm 11$ & $<0.33$ & $>0.12$\\
 	\hline    
    \end{tabular} \label{tab:comp}
 \end{table*} 
 
\subsection{J0906$+$6930}
 
J0906$+$6930 (also known as CGRaBS J0906$+$6930) was discovered as a high-redshift quasar from its optical spectroscopy by \citet{2004ApJ...610L...9R}. The redshift of $z = 5.47$ made the source the most distant blazar known at that time. It is also the most radio-luminous quasar at $z > 5$ \citep[radio loudness $S_{\rm 5GHz}/S_{\rm 0.44\mu m,rest} \sim 10^3$:][]{2004ApJ...610L...9R}. The source was detected in X-rays with the {\it Chandra} \citep{2006AJ....132.1959R} and {\it Swift} satellites \citep{2014ApJS..210....8E}. It was once suggested to be a promising $\gamma$-ray candidate source, however has not been detected by {\it Fermi} so far. 

The first VLBI image made with the VLBA by \citet{2004ApJ...610L...9R} at 15~GHz shows a compact core--jet structure extending to the southwest. The flux densities of the core and jet components are $115\pm 0.3$~mJy and $6.3\pm 0.4$~mJy, respectively. The separation between them is less than 1~mas. At 43~GHz, the compact core is detected with a flux density of $42 \pm 1.9$~mJy, but the weak jet was only marginally detected with a flux density of $4.1 \pm 1.1$~mJy (about 3.7 times the rms noise in the image).
We re-analyzed three epochs of archival VLBA 15-GHz data. The observational information is summarized in Table \ref{tab:obsVLBA}.

Figure \ref{fig:VLBImap} shows the emission structure of the source obtained from the KaVA and VLBA data. 
The KaVA image shown in Figure \ref{fig:VLBImap}-a only reveals a single unresolved core. 
A circular Gaussian brightness distribution model was used to fit the visibilities in {\sc Difmap} by using the task {\sc modelfit}.  
The derived core flux density is $74 \pm 11$~mJy, and the full-width-at-half-maximum (FWHM) size is 0.33~mas. 
The size is the same as obtained from 8-GHz VLBA data (Table~\ref{tab:comp}), but larger than the values from 15-GHz VLBA data, reflecting the fact that the core is not resolved by KaVA at 22~GHz, thus the measured core size is regarded as an upper limit.
The KaVA flux density is consistent with the interpolated value from the VLBA 15- and 43-GHz data.
In the VLBA image (Figure \ref{fig:VLBImap}-b), the jet extends in the southwestern direction in which the KaVA beam is more elongated and thus provides poorer resolution. Moreover, the jet flux density is at most 5~mJy at 22~GHz, as estimated from the 15-GHz flux density and its steep-spectrum nature. The expected jet emission is thus below 5 times of the KaVA image noise. The lower resolution of the KaVA image (1.9~mas\,$\times$\,0.9~mas) also works against detecting the expected jet extension at a separation of $\la1$~mas.

Figure \ref{fig:VLBImap}-b shows the image obtained from the VLBA archive data observed on 2005 May 15. It shows a core--jet structure, perfectly consistent with the previously published image in \citet{2004ApJ...610L...9R}. 
The new image has an even lower rms noise level than the 2004 February image due to the longer on-source time in the 2005 May observation.
The jet extends to southwest up to a distance  of about 1 mas. The jet peak is at about 0.7 mas southwest from the core.
Two circular Gaussian brightness distribution models were used to fit the visibilities. 
The fitted parameters from all VLBI data are shown in Table \ref{tab:comp}.
The core flux densities show good consistency at three epochs, indicating that there is no significant variability from 2004 to 2005. 
The jet flux densities show some difference, which could be related to the slightly different sensitivity and $(u,v)$ sampling of individual datasets.
We note that our fit for the 2004 data gives the core flux density of $132 \pm 9$ mJy, which is 17~mJy higher than the value obtained by \citet{2004ApJ...610L...9R}, but still within $2\sigma$ uncertainty. 

As discussed below and illustrated by Table \ref{tab:flux0906}, the source appeared to be in a quiescent state during the three epochs of the available VLBA 15 GHz data. These observations were conducted with nearly similar frequency setups allowing us to neglect possible frequency gradient in the source structure. VLBA images obtained in these three epochs do not show any profound structural features beyond strongly dominant core component. Finally, the time spread in the source's  rest frame is ($1 + z$) times shorter than at the observer's end, thus decreasing the time spread between the three epochs to just about two months. With all these in mind, we combined the three VLBA data sets for enhanced imaging. 
Indeed, the combined data have better $(u,v)$ coverage in both north--south and east--west directions than that presented in \citet{2004ApJ...610L...9R}, allowing us to restore the combined VLBA image with a high-resolution 0.6-mas circular beam (Figure~\ref{fig:VLBImap}-c). 
Benefiting from the good $(u,v)$ coverage, high sensitivity and high resolution of the combined 15-GHz VLBA image, we detected a weak feature S at about 1.5~mas south of the core, which was not seen previously.
In order to verify the new feature S, we separately created the VLBA image from the 2015 March 22 dataset which has a good ({\it u,v}) coverage in north--south direction, and also found S at the similar position although with a relatively lower signal-to-noise ratio when restoring the images with a 0.6 mas $\times$ 0.6 mas beam. 
The component S implies that the jet may have a sharp bending from the southwest to the south. Further high-sensitivity high-resolution images are needed to confirm this feature.

We checked the previously published VLA data on this source and found that they show good consistency with the VLBA data (Table \ref{tab:flux0906}). This suggests that there is no significant extended emission on arcsec scale. The 22-GHz flux density decreases from 2005 March to 2016 January, as seen from the KaVA and VLA data. These indicate either a slow variability over a time scale of years or a fraction of extended emission which is detected by the VLA but resolved by the VLBI. 

At 43 GHz, we did not detect clear fringes for J0906$+$6930 with KaVA. Therefore we can only set an upper limit of 57~mJy ($3\sigma$) for the flux density, which is consistent with the VLBA value reported by \citet{2004ApJ...610L...9R}. Similarly to the 15-GHz data, the 43-GHz VLBA and VLA flux densities show good agreement, supporting that there is no appreciable variability between epochs 2004 and 2005. 

\subsection{J0131$-$0321}
 
It was first discovered as a high-redshift quasar ($z =5.18 \pm 0.01$, SDSS J013127.34$-$032100.1) with the Lijiang 2.4 m and Magellan telescopes and identified as a radio-loud one \citep{2014ApJ...795L..29Y}. 
The source was claimed as a blazar based on {\it Swift} observations by \citet{2015MNRAS.450L..34G}. This classification was further strengthened by the inferred high brightness temperature from the European VLBI Network (EVN) observation at 1.7 GHz made by \citet{2015MNRAS.450L..57G}. This source was not detected in our KaVA experiment at 22 and 43~GHz. Note that the flux density at 1.7~GHz is 60~mJy, if the source has a flat or steep spectrum, the expected flux density at 22 GHz is $\lesssim60$ mJy, which is roughly the detection limit of the present 22 GHz KaVA data. There is also evidence for flux density variability \citep{2015MNRAS.450L..57G}. The 3-$\sigma$ upper limits for the flux densities in the compact radio structure derived from our KaVA non-detections are given in Table~\ref{tab:flux}.

\subsection{J1026$+$2542}
 
The source is also known as SDSS J102623.61$+$254259.5 and reported as a radio-loud AGN at $z = 5.266$ \citep{2012ApJS..203...21A}. It was assumed to be a blazar based on its strong and hard X-ray spectrum measured by {\it Swift} and the {\it Nuclear Spectroscopic Telescope Array (NuSTAR)} \citep{2012MNRAS.426L..91S,2013ApJ...777..147S}. The VLBI observations at 1.7 and 5 GHz show a prominent jet extending to a distance of $\sim$40~mas (corresponding to a projected linear size of $\sim$250~pc) \citep{2007ApJ...658..203H,2013MNRAS.431.1314F,2015MNRAS.446.2921F}. It is the first high-redshift ($z > 5$) quasar with measured apparent jet component proper motions which exceed $10\, c$ \citep{2015MNRAS.446.2921F}.  The radio spectrum index is $\alpha = -0.4$ \citep[][]{2013MNRAS.431.1314F,2017MNRAS.tmp..216C} and further steepens at frequencies above $\sim$30~GHz \citep[$\alpha = -0.7$,][]{2013ApJ...777..147S}. The core flux density at 5 GHz is $\sim$20 mJy,   resulting in an extrapolated flux density of $\sim$8 mJy at 22 GHz and $\sim$4 mJy at 43 GHz, respectively.
This source was not detected in our KaVA observations because of the insufficient sensitivity.

 \begin{table}
	\caption{Flux densities of J0906$+$6930 from interferometric measurements}
	\setlength{\tabcolsep}{4pt}
 \begin{tabular}{ccccc}
 	\hline
	$\nu$ & Date & Telescope & Flux density & Reference        \\
  	(GHz)	  &      &     & (mJy)      	&        \\ \hline
    1.4     & 1993 11 23 & VLA  & $93.4 \pm 2.8$& \cite{1998AJ....115.1693C} \\
    5       & 2005 03 04 & VLA  & $114 \pm 2.4$ & \cite{2006AJ....132.1959R} \\ 
    8.4     & 2005 03 04 & VLA  & $136 \pm 3.1$ & \cite{2006AJ....132.1959R} \\
            & 2011 06 03 & VLBA & $167 \pm 8$   & VLBA calibrator survey$^a$ \\ 
    15      & 2004 02 27 & VLBA & $139.4 \pm 13.0$   & \cite{2004ApJ...610L...9R}$^b$ \\    
            & 2005 03 04 & VLA  & $129 \pm 6.6$ & \cite{2006AJ....132.1959R} \\
            & 2005 03 22 & VLBA & $141.7 \pm 13.7$   & the present paper \\
            & 2005 05 15 & VLBA & $133.2 \pm 12.2$   & the present paper \\
    22      & 2005 03 04 & VLA  & $83 \pm 4.2 $ & \cite{2006AJ....132.1959R} \\
            & 2016 01 14 & KaVA & $74 \pm 11 $  & the present paper \\
    43      & 2004 03 03 & VLBA & $42 \pm 1.9 $ & \cite{2004ApJ...610L...9R} \\
            & 2005 03 04 & VLA  & $43 \pm 2.2 $ & \cite{2006AJ....132.1959R} \\
            & 2011 12 20 & KVN  & $108 \pm 13 $ & \cite{2012AJ....144..150P} \\
            & 2016 01 14 & KaVA & $<$57         & the present paper \\
	\hline
	\hline
 \end{tabular} \label{tab:flux0906}  \\ \vspace{1mm}
$^a$ data obtained from http://astrogeo.org maintained by L. Petrov \\
$^b$ flux density is from our own analysis
 \end{table} 
 
\section{Radio properties of J0906$+$6930} 
\label{sec4}

\subsection{Brightness temperature and Doppler boosting factor}

High apparent brightness temperature ($T_{\rm b}$) is an indicator of Doppler boosting of the relativistic jet in blazars \citep{1966ApJ...146..634P}. 
The maximum brightness temperature of an AGN core is constrained by the inverse Compton cooling to about $10^{11}$~K \citep{1969ApJ...155L..71K} for incoherent synchrotron emission by relativistic electrons when relativistic beaming is not taken into account. A lower value of $\sim5 \times 10^{10}$~K is derived if the particle energy density is in equipartition with the magnetic field \citep{1994ApJ...426...51R}. 
Observations of brightness temperatures higher than the equipartition $T_{\rm b}$ value are usually attributed to relativistic beaming. VLBI observations of a large AGN sample led to an intrinsic $T_{\rm b,int} \sim 2 \times 10^{11}$~K in the maximum brightness state \citep{2006ApJ...642L.115H}, exceeding the equipartition value. Even higher brightness temperature of $T_{\rm b} > 10^{13}$~K was detected in AO 0235$+$164 from the VLBI Space Observatory Programme (VSOP) observation \citep{2006PASJ...58..217F} and in two archetypal blazars, 3C\,273 \citep{2016ApJ...820L...9K} and BL\,Lac \citep{2016ApJ...817...96G} from the latest {\it RadioAstron} observations.

For the quasar J0906$+$6930, we calculated brightness temperature based on the KaVA and VLBA data (Table~\ref{tab:comp}). 
Because there are large fluctuations of the visibility amplitudes measured on the longest KaVA baselines, and the core is not resolved at the resolution of 1.9 mas $\times$ 0.9 mas, our size estimate obtained with Gaussian model fitting is taken as an upper limit and thus the lower limit of brightness temperature is set as 
 \begin{equation}
 T_{\rm b} = 1.22\times 10^{12}\frac{S}{\theta ^{2}\nu ^{2}}(1+z) \,\, {\rm K}, 
 \end{equation}\label{euq1}
where $S$ is the  flux density in Jy, $\theta$ the FWHM of the circular Gaussian component in mas, $z$ the redshift, and $\nu$ the observing frequency in GHz. The resulting brightness temperature is $> 1.2 \times 10^{10}$ K. 

The brightness temperature was also calculated using the VLBA data in the same way. The analysis of the 8.4-GHz data gave a higher value than that based on the KaVA data, $T_{\rm b} \sim 1.8 \times 10^{11}$~K. Three epochs of 15-GHz VLBA data obtained in 2004 and 2005 led to the estimates of $T_{\rm b}$ between (1.4--2.6)\,$\times\,10^{11}$~K. 
These values are comparable to the inverse Compton limit. The determined brightness temperatures indicate that the radio jet emission is beamed toward the observer. The Doppler boosting factor, $\delta = T_{\rm b} / T_{\rm b,int}$, is about 4, assuming that the intrinsic $T_{\rm b,int}$ is close to the equipartition value. In general, blazars have strongly beamed jet, high Doppler factor, and high brightness temperature. 
A study of a sample of the 173 brightest blazars shows that about one half of the sources have $T_{\rm b} > 10^{12}$~K, and the mean brightness temperature is $10^{12.0\pm0.8}$~K \citep{2011ApJ...742...27L}. The $T_{\rm b}$ of J0906$+$6930 places it at the low-end tail of the brightness temperature distribution, but it is still consistent with the general properties of these blazars.

\subsection{Spectrum and source classification}

\begin{figure}
	\centering
	\includegraphics[width=0.35\textwidth]{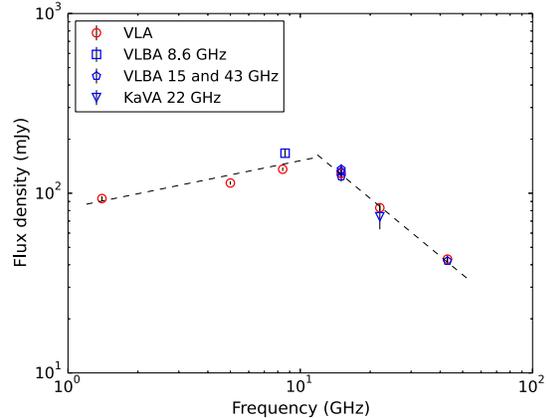} \vspace{5mm}
	\caption{Radio spectrum of J0906$+$6930. The data points are taken from Table~\ref{tab:flux0906}. The spectrum is divided into two sections: a flat-spectrum part from 1.4 to 8.4~GHz with a spectral index of $\alpha = 0.2$, and a steep-spectrum part from 15 to 43~GHz with a spectral index of $\alpha = -1.0$. The extrapolations of the two sections intersect at about~10 GHz.}
\label{fig:spec}
\end{figure}

\citet{2006AJ....132.1959R} observed the source with the VLA simultaneously at six frequencies from 1.4 to 43~GHz. They derived a spectrum with a changing slope: the lower-frequency section showed a flat or slightly rising spectrum with a spectral index of $\alpha^{8.4}_{1.4} = 0.21\pm 0.04$, and the higher-frequency section showed a steep spectrum with $\alpha^{43}_{15} = -1.00\pm 0.02$. The extrapolations of two sections intersect at around 10~GHz (Figure \ref{fig:spec}), corresponding to $\sim$65~GHz in the source rest frame. 
The radio properties, e.g., compact single-sided jet structure,  flat radio spectrum at GHz frequencies (in observer's frame) and high brightness temperature, are consistent with that the relativistic jet is beamed toward the observer. If it is indeed a flat-spectrum radio-loud quasar, the spectrum break at $\sim$65 GHz could be due to synchrotron ageing. 
The fit of multi-wavelength spectral energy distribution (SED) of J0906+6930 shows a bright X-ray Compton component, in agreement with a blazar identification of this source \citep{2004ApJ...610L...9R,2006AJ....132.1959R}. However the expected $\gamma$-ray emission was not detected by {\it Fermi} so far. 

\citet{2017MNRAS.tmp..216C} collected available flux density data of J0906$+$6930 from the literature from 148~MHz to 43~GHz, and the authors found that the flux densities at 148 and 325 MHz are far below the extrapolation of the flat-spectrum section to the low frequency end, implying there is significant absorption at the low frequencies with a turnover at about 0.45 GHz ($\sim$3 GHz in the source rest frame).
Low-frequency spectral turnover in AGN core might result from free-free absorption by ionized gas in the host galaxies; that usually happens at hundreds of MHz \citep[e.g.,][]{2005ApJ...634L..49A}.
The turnover at about 3 GHz due to free-free absorption in J0906$+$6930 cannot be ruled out at least due to the higher co-moving density of the interstellar medium that scales up proportionally to $(1+z^{3})$. 

On the other hand, \citet{2017MNRAS.tmp..216C} used a log parabola function to fit a convex radio spectrum with a peak frequency of $6.4 \pm 0.8$~GHz, which corresponds to $41.4 \pm 5.2$~GHz in the source rest frame (with the uncertainty likely underestimated). 
High-frequency peaked spectrum is often found in young radio galaxies and quasars, i.e.,  gigahertz peaked-spectrum (GPS) sources \citep[e.g.,][]{1991ApJ...380...66O,1998PASP..110..493O} and high-frequency peakers \citep[HFPs; e.g.][]{2000A&A...363..887D,2003PASA...20...79D}.
The compact source structure (a projected size of $\sim$5 pc), and peaked overall radio spectrum in J0906$+$6930 seem to be consistent with the properties of HFPs.
The peak frequency is correlated with source size \citep{2009AN....330..120F}; assuming a self-similar evolution of the extragalactic sources \citep[e.g.,][]{1996cyga.book..209B,1997MNRAS.286..215K,2012ApJ...760...77A}, the "peak frequency--source size" relation is converted to an observed anti-correlation between the rest-frame peak frequency and linear size (or age).  
The peak frequency of $\sim$40 GHz of J0906$+$6930 is by far the highest value. If it is indeed due to synchrotron self-absorption in a similar way with HFP sources, it suggests that J0906$+$6930 could be a newly-born radio AGN at an early cosmological epoch at $z = 5.47$, corresponding to 15 per cent of the present age of the Universe. Such a very young radio source which started its prominent radio activity in the recent past, likely within less than a few hundred years (measured in its rest frame) offers an excellent laboratory for studying radio AGN activity at their youngest age in the very early Universe.

More stringent constraints to the spectral fit and the peak frequency determination can be obtained by adding low-frequency measurements below 1~GHz with interferometers such as the Low-Frequency Array (LOFAR) and the Giant Metrewave Radio Telescope (GMRT). Following-up VLBI observations at multiple frequencies are also important for tracing the radio spectrum evolution to confirm or exclude its classification of a young radio quasar \citep[e.g.][]{2007A&A...475..813O,2008A&A...477..807O}.

\subsection{Variability}
\begin{figure}
	\centering
	\includegraphics[width=0.45\textwidth]{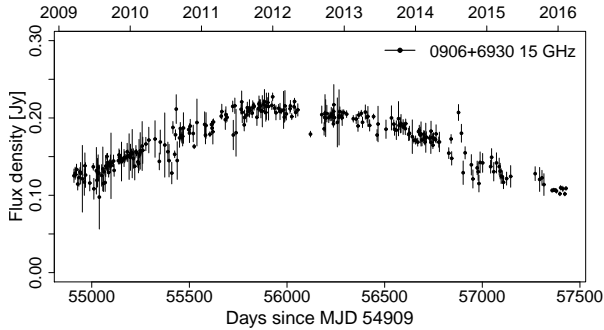}
	\caption{The light curve of J0906$+$6930 observed at 15~GHz with the OVRO 40-m radio telescope \citep{2011ApJS..194...29R}. The observations lasted from 2009 March 19 to 2016 February 10. 
}
	\label{fig:lc}
\end{figure}

Figure~\ref{fig:lc} shows the light curve measured at 15~GHz with the Owens Valley Radio Observatory 40-m radio telescope \citep{2011ApJS..194...29R}. The  observations lasted from 2009 March 19 to 2016 February 10. During the 7-year monitoring period, the source shows slow long-term variability. The variability index is : 
\begin{equation}
V = \frac{S_{\rm max} - S_{\rm min}}{S_{\rm max} + S_{\rm min}},
\end{equation}
where $S_{\rm min}$ and $S_{\rm max}$ are the minimum and maximum flux densities, respectively. 
The derived variability index for J0906$+$6930, $V = 0.3$, is significantly lower than that of blazars \citep[e.g.,][]{2008A&A...485...51H}. 
However, we should mention that the 7-year total flux density monitoring period corresponds to a period shorter by a factor of $(1+z)$, i.e., just over 1 year, in the rest frame because of the cosmological time dilation. We note that the source flux density was close to its minimum at the time of our KaVA observations in 2016 January, contributing to the difficulty of the detection.

Table~\ref{tab:flux0906} lists interferometric measurements of the flux densities collected from the literature and the present work.
The core flux densities obtained from the VLA and VLBA at the same frequencies (15 and 43~GHz), though they were not measured at the same epochs, show good consistency. This indicates that the compact core dominates the total flux density, and that the source did not have significant variability from 2004 February to 2005 March, as is also suggested by the slowly-varying 15-GHz light curve in Fig.~\ref{fig:lc}. The KaVA flux density is also consistent with the VLA value at 22~GHz within the uncertainties, implying no significant difference between the two widely-spaced epochs of 2005 March 4 and 2016 January 14. 
The 8.4-GHz flux density observed with the VLBA on 2011 June 3 is not much different from the VLA measurement on 2005 March 4 either. All these pieces of evidence lead us to conclude that the compact radio structure does not have violent flux density variability. 

The 43-GHz observations made by the KVN in 2011 December in the QCAL-1 observing campaign inferred a flux density of $108 \pm 13$\,mJy \citep{2012AJ....144..150P}, which is 2.5 times higher than the values in 2004 and 2005 (Table~\ref{tab:flux0906}). We mention that, at least at 15~GHz, the radio flux density of the source was at its maximum at the end of 2011 (Fig.~\ref{fig:lc}) when the QCAL-1  campaign was carried out. The ratio of the maximum to minimum flux densities at 15~GHz is about 2, similar to the factor of 2.5 at 43~GHz. Therefore it could be that the source was at a quiescent state in 2004 and 2005, while it flared at the end of 2011.
 
For a Doppler factor $\delta \sim 4$, the time elapsed in the source frame is $\delta/(1+z) \sim 0.62 $ times that in the observer frame $\Delta t \sim 4.3$ yr corresponding to a compact region of size $c \Delta t \sim 1.3$ pc. If this region is along the jet axis at kpc-scale distances from the BH, it can correspond to a typical timescale over which radiation pressure due to AGN activity balances or alters the large-scale accretion \cite[e.g.][]{2017arXiv170106565P} as the radius at which the SMBH has a gravitational influence on the accreting gas, roughly given by the Bondi radius \cite[e.g.][]{2002apa..book.....F} $R_B = G M_\bullet/c^2_s \sim 260$ kpc for a central SMBH mass of $3 \times 10^9 M_\odot$ \citep{2015MNRAS.450L..34G} and fiducial local sound speed of 10 km s$^{-1}$ for a gas at $10^4$ K. This will be accompanied by a slow variability. On the other hand, if this region is at pc scales, the plausible sequence of events involves the brightening of the radio core due to a shock propagating outward accompanied by the possible ejection of a jet component \cite[e.g.][]{2010ApJ...710L.126M}. In this case, variability will be slow if the shock propagates slowly owing to low injection rate into the jet or rapid interaction of the jet with the surrounding interstellar medium and the morphology will be compact when observed in snapshots. Such a scenario is expected in the case of young AGN with low brightness temperatures and consequent low Doppler factors which in addition is diffuse in gamma-ray emission as in the case of 3C 286 \cite[][]{2017MNRAS.466..952A}. It is thus essential for additional observations to distinguish between these distinct scenarios.

\section{summary}
\label{sec5}

Three radio-loud quasars at the highest redshifts ($z > 5$) were observed with the KVN and VERA Array (KaVA) at 22 and 43~GHz,  to demonstrate its capability to image extragalactic radio sources. J0906$+$6930 was detected at 22 GHz, and the image shows an unresolved core. This object was not detected at 43~GHz, like two other sources (J0131$-$0321 and J1026$+$2542) at either 22 and 43 GHz, because of the insufficient sensitivity  of the present observations and the low flux density resulting from the steep spectrum of the core. 
Four epochs of archival VLBA data of J0906$+$6930, three at 15~GHz and one at 8.4~GHz, were also imaged and re-analyzed. 
Owing to the good ($u, v$) coverage and long integration time, the new images are superior to the previously published one, and have higher sensitivity and higher resolution.
A new feature is detected to the south of the core, likely implying a sharp jet bending from southwest to south.
The compact source structure, the flat spectrum at tens GHz frequencies (in source rest frame) and the high brightness temperature of the core reinforce that J0906$+$6930 is a blazar and the jet is beamed toward the observer.  
As extended jets are rarely detected in $z\geq5$ blazars so far, J0906+6930 holds strong potential for follow-up VLBI imaging to detect proper motion in this extremely high-$z$ quasar. 
For the first time, long-term single-dish monitoring light curve was presented for such a high-$z$ radio-loud quasar. 
The variability might be related to the processes in the jet.

So far, little information is available on the long-term morphological and flux density behaviour of these distant sources. Follow-up and monitoring observations are essential to better understand the cosmological evolution of galaxies, AGN activity triggers, and to populate the angular size--redshift and proper motion--redshift relations with measured data for tests of cosmological expansion. Elucidating the role of extremely high-redshift sources in the luminosity function is useful for future observations and surveys, and for understanding their relation to radio-loud AGN in the nearby universe.

\section*{Acknowledgements}

The authors are supported in this work by SKA pre-construction funding from the China Ministry of Science and Technology (MOST, No 2016YFE0100300) and the Chinese Academy of Sciences (CAS), the China--Hungary Collaboration and Exchange Programme by the International Cooperation Bureau of the CAS.
TA thanks the grant supported by the Youth Innovation Promotion Association of CAS. 
SF and K\'EG thank for the Hungarian National Research, Development and Innovation Office (OTKA NN110333) for support.
PM thanks the CAS President's International Fellowship Initiative (CAS-PIFI; grant no. 2016PM024) post-doctoral fellowship and the NSFC Research Fund for International Young Scientists (grant no. 11650110438). 
The National Radio Astronomy Observatory is a facility of the National Science Foundation operated under cooperative agreement by Associated Universities, Inc. 
We acknowledge the use of Astrogeo Center database of brightness distributions, correlated flux densities, and images of compact radio sources produced with VLBI.
This research has made use of data from the OVRO 40-m monitoring program \citep{2011ApJS..194...29R} which is supported in part by NASA grants NNX08AW31G, NNX11A043G, and NNX14AQ89G, and NSF grants AST-0808050 and AST-1109911.

\end{document}